\documentclass[twocolumn]{revtex4}
\usepackage{amsfonts}
\usepackage{amsmath}

\begin{document}

\title{Energy average formula of photon gas rederived by using the
generalized Hermann-Feynman theorem}
\author{Hong-yi Fan }
\affiliation{Department of Physics, Shanghai Jiao Tong University, Shanghai, 200030, Chin}

\begin{abstract}
By virtue of the generalized Hermann-Feynmam theorem and the method of
characteristics we rederive energy average formula of photon gas, this is
another useful application of the theorem.


key words: generalized Hermann-Feynman theorem, Energy average formula
\end{abstract}

\maketitle

\section{Introduction}

It is common knowledge in the black-body radiation that the expectation
value of a Planck oscillator of frequency $\varpi_{s}$ is (excluding the
zero-point\ energy term) is%
\begin{equation}
\bar{E}_{s}=\frac{\hbar \varpi_{s}}{e^{\beta \hbar \varpi_{s}}-1}   \label{1}
\end{equation}
Historically, Planck derived this result in 1900 first by using
thermodynamics and then by classical statistics. Later in 1924 Bose studied
question such as the "probability of an energy level $E_{s}$ ($=\hbar
\varpi_{s}$ $)$ being occupied by $n_{s}$ photons at a time", he found the
mean value of $n_{s}$ is $\frac{1}{e^{\beta \hbar \varpi_{s}}-1}.$ Einstein,
after reading Bose's paper, concluded that the basic fact to remember during
the process of distributing the various photons over the various energy
levels is that the photons are indistinguishable---a fact that had been
implicitly taken care of in Bose's treatment (he essentially treated the
light quanta as particles of a gas). Planck and Bose-Einstein obtain the
same mean photon number and mean energy formlas with different approaches.
We then ask: is there any other method with which we can rederive this
formula? If yes, then this method is worth paying attention in quantum
statistics. The purpose of this work is to re-derive the energy average
formula of photon gas by using the generalized Hermann-Feynman theorem, a
theorem regarding to the ensemble average of mixed states.

In quantum mechanics the usual Hermann-Feynman theorem\cite{1,2} is for pure
states, which states that%
\begin{equation}
\frac{\partial E_{j}\left( \chi \right) }{\partial \chi}=\left \langle
j\right \vert \frac{\partial H\left( \chi \right) }{\partial \chi}\left
\vert j\right \rangle ,\text{ \ \ }\left \langle j\right. \left \vert
j\right \rangle =1,   \label{2}
\end{equation}
where $H$ is the Hamiltonian depending on the parameter $\chi$, and$\ E_{j}$
and $\left \vert j\right \rangle $ are energy eigenvalues and eigenvectors
of $H$, respectively. This theorem has been widely used in molecular
physics, quantum chemistry and quantum statistics. Having noticed that Eq.(%
\ref{1}) is just for pure state expectation value, we need a generalized
Hellmann-Feynman theorem for mixed state in the sense of ensemble average in
quantum statistics. In Ref.\cite{3} the generalized Hellmann-Feynman theorem
(GHFT) has been set up
\begin{equation}
\frac{\partial}{\partial \chi}\left \langle H\right \rangle _{e}=\left
\langle \left[ 1+\beta \left \langle H\right \rangle _{e}-\beta H\right]
\frac{\partial H}{\partial \chi}\right \rangle _{e},   \label{3}
\end{equation}
where the subscript $e$ stands for the ensemble average, $\beta=\left(
kT\right) ^{-1}$, $k$ is the Boltzmann constant, and $\left \langle A\right
\rangle _{e}\equiv Tr\left( e^{-\beta H}A\right) /Tr\left( e^{-\beta
H}\right) $ for arbitrary operator $A$. When $H$ is $\beta$ independent, $%
\frac{\partial H}{\partial \beta}=0,$ Eq.(\ref{3}) can be reformed as
\begin{equation}
\frac{\partial}{\partial \chi}\left \langle H\right \rangle _{e}=\frac{%
\partial }{\partial \beta}\left[ \beta \left \langle \frac{\partial H}{%
\partial \chi }\right \rangle _{e}\right] .   \label{4}
\end{equation}
Now for the Bose oscillator Hamiltonian $H=\hbar \varpi a^{\dagger}a,$ using
(\ref{4}) we see
\begin{align}
\frac{\partial}{\partial \varpi}\left \langle H\right \rangle _{e} & =\frac{%
\partial}{\partial \beta}\left[ \beta \left \langle \frac{\partial H}{%
\partial \varpi}\right \rangle _{e}\right] =\frac{\partial}{\partial \beta }%
\left[ \beta \left \langle \hbar a^{\dagger}a\right \rangle _{e}\right]
\notag \\
& =\frac{\partial}{\varpi \partial \beta}\left[ \beta \left \langle \hbar
\varpi a^{\dagger}a\right \rangle _{e}\right] =\frac{\partial}{\varpi
\partial \beta }\left[ \beta \left \langle H\right \rangle _{e}\right]
\notag \\
& =\frac{1}{\varpi}\left \langle H\right \rangle _{e}+\frac{\beta \partial }{%
\varpi \partial \beta}\left[ \left \langle H\right \rangle _{e}\right] ,
\label{5}
\end{align}
that is%
\begin{equation}
\left( \varpi \frac{\partial}{\partial \varpi}-\beta \frac{\partial}{%
\partial \beta}\left[ \left \langle H\right \rangle _{e}\right] \right)
\left \langle H\right \rangle _{e}=\left \langle H\right \rangle _{e},
\label{6}
\end{equation}
which can be solved by virtue of the method of characteristics \cite{4},
according to it we have the equation

\begin{equation}
\frac{d\varpi}{\varpi}=-\frac{d\beta}{\beta}=\frac{d\left \langle H\right
\rangle _{e}}{\left \langle H\right \rangle _{e}}.   \label{7}
\end{equation}
It then follows two integration constants
\begin{equation}
\ln \varpi=-\ln \beta+C_{1},\text{ }\ln \left \langle H\right \rangle
_{e}-\ln \varpi=C_{2},   \label{8}
\end{equation}
which tells us that $\left \langle H\right \rangle _{e}$ is of the form%
\begin{equation}
\left \langle H\right \rangle _{e}=\varpi F\left( \beta \varpi \right)
\label{9}
\end{equation}
the form of $F$ is to be determined. Eq. (\ref{9}) can be confirmed by
noticing%
\begin{equation}
\varpi \frac{\partial}{\partial \varpi}=\frac{\partial}{\partial \ln \varpi }%
,\text{ }\ln \varpi \equiv x,\text{ }\ln \beta \equiv y,   \label{10}
\end{equation}
and rewriting (\ref{6}) as
\begin{equation}
\left( \frac{\partial}{\partial x}-1\right) \left \langle H\right \rangle
_{e}=\frac{\partial}{\partial y}\left[ \left \langle H\right \rangle _{e}%
\right] ,   \label{11}
\end{equation}
its solution is%
\begin{equation}
\left \langle H\right \rangle _{e}=e^{x}g\left( x+y\right) =\varpi F\left(
\beta \varpi \right) .   \label{12}
\end{equation}
In order to make sure of $F^{\prime}$s concrete form$,$ we introduce $%
a=\left( \sqrt{\frac{m\varpi}{\hbar}}X+i\frac{P}{\sqrt{m\varpi \hbar}}%
P\right) /\sqrt{2},$ so up to a zero-point energy term,
\begin{equation}
H=\frac{P^{2}}{2m}+\frac{1}{2}m\varpi^{2}X^{2}.   \label{13}
\end{equation}
We see when $\varpi \rightarrow0,$ $H\rightarrow \frac{p^{2}}{2m}$ (free
particle), Eq. (\ref{9}) reduces to%
\begin{equation}
\lim_{\varpi \rightarrow0}\left \langle H\right \rangle _{e}=\lim_{\varpi
\rightarrow0}\varpi F\left( \beta \varpi \right) ,   \label{14}
\end{equation}
so if $\lim_{\varpi \rightarrow0}\varpi F\left( \beta \varpi \right) $ is
finite,%
\begin{equation}
\lim_{\varpi \rightarrow0}F\left( \beta \varpi \right) \rightarrow \infty.
\label{15}
\end{equation}
\ On the other hand, we recall the Bloch equation for the density matrix $%
\rho$ of a free particle
\begin{equation}
-\frac{\partial \rho \left( p,\beta \right) }{\partial \beta}=\frac{p^{2}}{2m%
}\rho \left( p,\beta \right) ,   \label{16}
\end{equation}
its normalized solution
\begin{equation}
\rho \left( p,\beta \right) =\sqrt{\frac{\beta}{2m\pi}}e^{-\beta
p^{2}/\left( 2m\right) },\text{ \ }Tr\rho=1.   \label{17}
\end{equation}
so the energy average for $H_{f}=\frac{P^{2}}{2m}$ is
\begin{align}
\left \langle H_{f}\right \rangle _{e} & \equiv \left \langle \frac{P^{2}}{2m%
}\right \rangle _{e}=\sqrt{\frac{\beta}{2m\pi}}\int_{-\infty}^{\infty}\frac{%
p^{2}}{2m}e^{-\beta p^{2}/\left( 2m\right) }dp  \notag \\
& =\sqrt{\frac{\beta}{2m\pi}}\left( -\frac{\partial}{\partial \beta}\right)
\int_{-\infty}^{\infty}e^{-\beta p^{2}/\left( 2m\right) }dp  \notag \\
& =\frac{1}{2\beta}=\frac{KT}{2}.   \label{18}
\end{align}
Combining (\ref{18}) and (\ref{14}) we see
\begin{equation}
\left \langle H_{f}\right \rangle _{e}=\lim_{\varpi \rightarrow0}\left
\langle H\right \rangle _{e}=\lim_{\varpi \rightarrow0}\varpi F\left( \beta
\varpi \right) =0\times \infty=\frac{1}{2\beta}.   \label{19}
\end{equation}
Using the rule of searching for limitation,
\begin{equation}
\frac{1}{\beta}=\lim_{\varpi \rightarrow0}\frac{\hbar}{\beta \hbar e^{\beta
\hbar \varpi}}=\hbar \lim_{\varpi \rightarrow0}\varpi \times \frac{1}{%
e^{\beta \hbar \varpi}-1},   \label{20}
\end{equation}
comparing \ref{20}) with (\ref{19}) and considering (\ref{15}) we can
determine the form of $F\left( \beta \varpi \right) ,$
\begin{equation}
F\left( \beta \varpi \right) =\frac{1}{e^{\beta \hbar \varpi}-1},\text{\ }
\label{21}
\end{equation}
so follows $\left \langle H\right \rangle _{e}=\frac{\hbar \varpi}{e^{\beta
\hbar \varpi}-1}.$

Besides, using the GHFT we\ can also provide:

1. average kinetic energy $=$ average potential energy. To\ prove this, we
notice

\begin{equation}
\left[ XP,H\right] =i\frac{P^{2}}{m}-im\varpi^{2}X^{2},   \label{22}
\end{equation}
due to%
\begin{equation}
0=\left \langle \left[ XP,H\right] \right \rangle _{e}=i\left \langle \frac{%
P^{2}}{m}\right \rangle _{e}-i\left \langle m\varpi^{2}X^{2}\right \rangle
_{e}   \label{23}
\end{equation}
so%
\begin{align}
\left \langle H\right \rangle _{e} & =\left \langle \frac{p^{2}}{2m}+\frac {1%
}{2}m\varpi^{2}x^{2}\right \rangle _{e}  \notag \\
& =\left \langle \frac{p^{2}}{m}\right \rangle _{e}=\left \langle m\varpi
^{2}x^{2}\right \rangle _{e}.   \label{24}
\end{align}

2. Though $H$ $\ $involves the mass, $\left \langle H\right \rangle _{e}$ is
$m$ independent. In fact, using (23) and (2) we have%
\begin{align}
\frac{\partial}{\partial m}\left \langle H\right \rangle _{e} & =\frac {%
\partial}{\partial \beta}\left[ \beta \left \langle \frac{\partial H}{%
\partial m}\right \rangle _{e}\right]  \notag \\
& =\frac{\partial}{\partial \beta}\left \langle \beta \left( \frac{\varpi^{2}%
}{2}x^{2}-\frac{1}{2m^{2}}p^{2}\right) \right \rangle _{e}=0.   \label{25}
\end{align}

\ In sum, we have rederived the energy average formula of photon gas (Bose
distribution) by using the GHFT, this is another useful application of the
theorem. The other application can be seen in Ref.\cite{5}.

\end{document}